\documentclass [namedreferences]{kluwer}    
\usepackage{graphicx}
\newdisplay{guess}{Conjecture}
\def\eqd{\,{\buildrel d \over =}\,}
\begin{document}
\begin{article}
\begin{opening}
\title{ Toward synthesis of  solar wind and geomagnetic scaling exponents: a fractional
L\'{e}vy motion model}
\author{Nicholas W. \surname{Watkins}$^1$\email{NWW@bas.ac.uk}}
\author{Daniel \surname{Credgington}$^1$}
\author{Bogdan \surname{Hnat}$^2$}
\author{Sandra C. \surname{Chapman}$^2$}
\author{Mervyn P. \surname{Freeman}$^1$}
\author{John \surname{Greenhough}$^3$}
\runningauthor{N. W. Watkins et al.}
 \runningtitle{Synthesis of scaling exponents: a fractional L\'{e}vy model}
 \institute{$^1$ British Antarctic Survey, High Cross, Madingley Road, Cambridge,
 CB3 0ET, UK} \institute{$^2$ Space and Astrophysics Group, University of Warwick, Coventry CV4 7AL, UK}
\institute{$^3$ Space Research Centre, University of Leicester,
Leicester LE1 7RH, UK}
\date{October 27, 2005}

\begin{abstract}
Mandelbrot introduced the concept of fractals to describe the
non-Euclidean shape of many aspects of the natural world. In the
time series context he proposed the use of fractional Brownian
motion (fBm) to model non-negligible temporal persistence, the
``Joseph Effect"; and L\'{e}vy flights to quantify large
discontinuities, the ``Noah Effect". In space physics, both
effects are manifested in the intermittency and long-range
correlation which are by now well-established features of
geomagnetic indices and their solar wind drivers. In order to
capture and quantify the Noah and Joseph effects in one compact
model we propose the application of the ``bridging" fractional
L\'{e}vy motion (fLm) to space physics. We perform an initial
evaluation of some previous scaling results in this paradigm, and
show how fLm can model the previously observed exponents. We
suggest some new directions for the future.

\end{abstract}

\end{opening}
\vspace{-.25in}
\section{Introduction}
Ever since it became clear that Earth's magnetosphere is
influenced by the sun, significant effort has been devoted to
establishing the relationship between fluctuations in the energy
delivered by the solar wind to the magnetosphere and variations in
the magnetospheric response. A particularly important
diagnostic for the response has been the family of geomagnetic
indices, especially the Auroral Electrojet index $AE$
\cite{davis}. A common proxy for the solar wind input is the
$\varepsilon$ function \cite{perrault} which estimates the fraction
of the solar wind Poynting flux through the dayside magnetosphere.

One approach is to investigate causal relationships, and
considerable sophistication has now been developed in this (e.g.
\citeauthor{ukhorskiy2004},\citeyear{ukhorskiy2004};\citeauthor{march2005},\citeyear{march2005}
and references therein). However, even without examining
causality, significant information can be obtained by examining
the scaling behaviour of fluctuations. A first analysis of this,
in the Fourier domain, was done by
\citeauthor{tsurutani}\shortcite{tsurutani} using the power
spectrum. Subsequent analyses have introduced other methods for
detecting scale invariance
(e.g. \citeauthor{TakaloGRL},\citeyear{TakaloGRL};\citeauthor{freeman2000a},\citeyear{freeman2000a,freeman2000b};
\citeauthor{uritsky},\citeyear{uritsky}). Most recently
\citeauthor{hnat2002a},
\shortcite{hnat2002a,hnat2002b,hnat2003a,hnat2003b,hnat2005} and
\citeauthor{chapman2005},\shortcite{chapman2005} have studied the
scaling collapse of the increments of time series.

 A  fundamental problem has   been raised by the evidence for multifractality in  some solar
 wind quantities (e. g. \citeauthor{hnat2002a},\shortcite{hnat2002a} and references therein) and the $AE$
index \cite{consolini96}. Multifractality is physically well
motivated-at least for solar wind quantities-in that it arises
naturally from the intermittency of  multiplicative turbulent
cascade models \cite{frisch}. Multifractality would imply that the
Hurst's ``roughness" exponent $H$ is not constant but varies from
scale to scale. This evidence for multifractality in the indices
thus means that any comparison of pairs of scaling exponents
derived from solar wind and geomagnetic indices may be problematic
\cite{WatkinsNPG,chang2001}. Preliminary comparisons of solar wind
and geomagnetic field measurements  made using multiscaling
measures \cite{voros} showed similarity at low orders after low
pass filtering of the magnetospheric quantities. However,
\citeauthor{hnat2002a}(2002-2005),
 in examining a range of solar wind quantities, have recently found some
apparent simplifications. They see the intriguing result that
although some quantities (notably $v$ and $B$) {\it do not} show a
simple scaling collapse, consistent with their well-known
multifractality, others (such as $B^2$) {\it do} i.e. they are, in
this sense, effectively monofractal. Recently
\citeauthor{hnat2003b}\shortcite{hnat2003b} have extended the 1
year $AE/U/L$ dataset studied by
\citeauthor{hnat2002b}\shortcite{hnat2002b}to the 10 years used by
\citeauthor{freeman2000a}\shortcite{freeman2000a}. They find that
when such long auroral index datasets are examined, $AE$ and
$\varepsilon$ do indeed have discernably different PDFs.

Such analyses are not easy to compare. Some used overlapping index
and solar wind time series \cite{uritsky}, other did not
\cite{freeman2000a}. Techniques which impose finite limits on the
integral used to evaluate structure functions have also been
explored (\cite{chapman2005} and references therein). The choice
of solar wind measures and geomagnetic time series has also
varied. It seems to us  thus imperative to try to start to
reconcile the various studies and understand why some show much
greater similarity between the solar wind signal and indices than
others. We also believe that the synthesis of observations will
help towards a goal we have proposed elsewhere: The definition of
models which are either I) simple, statistical, ``strawman" models
which may nonetheless capture some relevant fluctuation
phenomenology e.g. the fractional lognormal model sketched by
\citeauthor{WatkinsNPG}\shortcite{WatkinsNPG}) or II) more clearly
statistical physics-based e.g. the generalised Fokker-Planck model
of \citeauthor{hnat2005}\shortcite{hnat2005} and
\citeauthor{chapman2005}\shortcite{chapman2005}).

By analogy with mathematical economics we may think of the Type I
models  as modelling the ``stylized facts" of the coupled solar
wind magnetospheric system \cite{WatkinsNPG}. In this paper we
shall introduce one such model: fractional L\'{e}vy motion
\cite{mandelbrot1995,chechkin2}, in order to see how well it can
describe the solar wind  $\varepsilon$ function and the $AE$ family
of indices ($AE$ itself, $AU$ and $AL$). Preliminary comparison is
made with some of the
 measurements listed above, and it is shown that the model
 provides a good quantitative explanation for the difference
 between two scaling exponents first noted in this context  by
 \citeauthor{hnat2002a}\shortcite{hnat2002a}
  as well as a possible qualitative explanation for the multifractal behaviour seen
by \citeauthor{hnat2003b}\shortcite{hnat2003b}. Where relevant, the effect of the truncation (finite variance) implicit in both a natural data series and a computer model are noted. Future directions
are then sketched.

\section{Datasets used}

The $AE$ and $\varepsilon$ data are a 1 year subset of those studied
by \citeauthor{hnat2002b}\shortcite{hnat2002b,hnat2003a}. They
correspond to the years 1978 and 1995 respectively. As with \cite{hnat2002b} solar wind data is
taken only for periods when WIND is definitely in the solar wind, see \cite{freeman2000a} for details.
We follow \citeauthor{hnat2002b}\shortcite{hnat2002b} by firstly
differencing the time series $X(t)$ of the indices $AE,AU, AL$ and
$\varepsilon$ at intervals $\tau$ of $1,2,3 \ldots$ times the
fundamental sampling period (1 minute for the indices and and 46
seconds for $\varepsilon$) to generate difference time series $\delta
X(t,\tau)=X(t+\tau)-X(t)$. For further details of the dataset and
preprocessing techniques see \cite{hnat2002b} and references
therein.

\vspace{-.3in}

\section{Motivation for and testing of a fractional L\'{e}vy motion model}

\subsection{Fractional L\'{e}vy motion as a bridge between L\'{e}vy flights and  fractional Brownian motion}

As noted by \citeauthor{mandelbrot1995}\shortcite{mandelbrot1995}:
\begin{quote} The ``normal"
model of natural fluctuations is the Wiener Brownian motion
process (WBm). By this standard, however, many natural
fluctuations exhibit clear-cut ``anomalies" which may be due to
large discontinuities (``Noah Effect") and/or non-negligible
global statistical dependence (``Joseph Effect"). [Mandelbrot's
book ``The Fractal Geometry of Nature"] ... shows that one can
model various instances of the Noah effect by the classical
process of [standard L\'{e}vy motion] (sLm), and various instances
of the Joseph effect by the process of [fractional Brownian
motion] (fBm).
\end{quote}
 \citeauthor{TakaloGRL}\shortcite{TakaloGRL} were the first to use fBm  as a model of
the auroral indices, but it subsequently could not describe the
highly non-Gaussian leptokurtic distributions seen in differenced
solar wind  and geomagnetic index quantities. This can for example
be seen in Fig. 7 of \cite{chapman2005} where the pdf of
differences $\delta X$ of $AE$ is contrasted with the Gaussian pdf
of an fBm with equal Hurst exponent $H$. Similarly we are are
aware of only a small number
\cite{kabin1998,consolini1997b,hnat2002a,bruno2004} of discussions
of the use of truncated sLm as a model for in-situ solar wind,
magnetotail or ground-based magnetometer time series. One reason
why sLm has not seen wider use here is because it cannot reproduce
the correlated increments seen for both these types of data and
also because it models superdiffusive ($H>0.5$) rather than the
observed subdiffusive ($H<0.5$) behaviour. The term ``truncated
L\'{e}vy flight" usually indicates standard L\'{e}vy motion with a
finite variance introduced deliberately by means of a finite range
cutoff (c.f. the discussion in section 8.4 of
\citeauthor{MantegnaStanley2000}\shortcite{MantegnaStanley2000});
however any finite (and thus finite-variance) series of sLm must also be
naturally truncated, but in an uncontrolled fashion \cite{nakao}.

\citeauthor{mandelbrot1995}\shortcite{mandelbrot1995} went on to
note that:
\begin{quote} sLm and fBm, however, are far from exhausting the
anomalies found in nature ...  many phenomena exhibit {\em both}
the Noah and Joseph effects and fail to be represented by either
sLm or fBm ... One obvious bridge, fractional L\'{e}vy motion, is
interesting mathematically, but has found no concrete use".
\end{quote}
Since those words were written, fLm has found  applications,
notably in geophysics \cite{painter} and telecommunications
network modelling \cite{laskin2002}. We here apply it to
essentially the same need; to compactly describe and unify the
``stylized facts" of the well-demonstrated Noah and Joseph effects
in space plasma physics time series \cite{WatkinsNPG}.

\subsection{Mathematical definition of fractional L\'{e}vy motion}

Fractional L\'{e}vy motion  can be defined using a
Riemann-Liouville fractional integral  generalising the
better-known expression for fractional Brownian motion
\cite{voss1985}. We here adapt the notation of equation 5 of
\citeauthor{laskin2002}\shortcite{laskin2002}, which defines a
process $W_{\mu,\beta}$:

\begin{equation}
W_{\mu,\beta}(t) = \frac{1}{\Gamma (\beta/2)}\int_{0}^{t}
(t-\tau)^{(\beta/2-1)}dW_{\mu}(\tau) \label{eq:fLm}
\end{equation}

Equation (\ref{eq:fLm}) can be unpacked  as a summation of
L\'{e}vy stable increments $dW_{\mu}(\tau)$ each  weighted by a
response function $(t-\tau)^{(\beta/2-1)}$. The $\mu$ parameter
describes the power law tail of the pdf of $dW$ which falls off as
$P(x) \sim x^{-(1+\mu)}$. $\mu=2$ is the special, Gaussian, case
corresponding to fBm. $\beta$ is the parameter which controls
long-range dependence. It is well known to be related to the power
spectral density $S(f)\sim f^{-\beta}$ for fractal processes with
finite variance \cite{voss1985}, but can also be rigorously
defined through fractional differentiation in other cases
\cite{chechkin2}.

With $\mu=2$ and taking in addition $\beta=2$  the response
function becomes unity, giving an uncorrelated random Gaussian
walk (WBm). Keeping $\beta=2$ but allowing $\mu$ to vary in the
range $0$ to $2$ describes sLm. fLm is thus in general a process
with $\mu,\beta$ allowed to vary in the range $[0 < \mu \le 2, 1
\le \beta \le 3]$ and so forms a bridge between the $\beta=2$ sLm
and $\mu=2$ fBm ``axes". fLm thus by construction exhibits both
the sources of anomalous diffusion identified by Mandelbrot above.

These limits have corresponding simplified Fractional Kinetic
Equations (FKE) for the pdf $P(W)$, see section 5.2 of
\cite{zaslavsky}. Putting $W=W_{\mu,\beta'}(x,t)$ with
$\beta'=\beta/2$, WBm is given by the diffusion equation
$\partial^1_t P(W_{2,1})=\partial_x^2 (\mathcal{A} P(W_{2,1}))$;
fBm by $\partial^{\beta'}_t P(W_{2,\beta'})=\partial_x^2
(\mathcal{A} P(W_{2,\beta'}))$; and sLm by $\partial^1_t
P(W_{\mu,2})=\partial_x^{\mu} (\mathcal{A} P(W_{\mu,2}))$. fLm
should thus correspond to equation (132) of \cite{zaslavsky}:
\begin{equation}
\frac{\partial^{\beta'}}{\partial t^{\beta'}}P(W_{\mu,\beta'}) =
\frac{\partial^{\mu}}{\partial |x|^{\mu} } ( \mathcal{A}
P(W_{\mu,\beta'}))\label{eq:FKE}
\end{equation}
All cases have a fixed diffusion constant $\mathcal{A}$.
Future work is required to establish if this simplified form of
equation (127) of \cite{zaslavsky}, the full FKE, can map on to
the Fokker-Planck equation of \cite{hnat2005} or whether the full
equation, including fractional drift and diffusion terms, is
needed. After initial submission of this paper we also became aware of the relevance of the
work of \citeauthor{milovanov2001}\shortcite{milovanov2001} to the interpretation of fLm as an FKE; 
see in particular their equation (3). 

\subsection{Self-similarity, the Hurst exponent and peak scaling}

We now follow \citeauthor{laskin2002}\shortcite{laskin2002} to
show  that $W_{\mu,\beta}$ is indeed an H-selfsimilar process.
 To see this  we first put $\tau=cs$ in (\ref{eq:fLm}).
We then use the fact that the increments $dW_{\mu}(cs)$ are
defined to be $1/\mu$ self-similar i.e. are equal in distribution
($\eqd$) to $c^{1/\mu} dW_{\mu}(s)$. Then
\begin{equation}
W_{\mu,\beta}(ct) \eqd c^H W_{\mu,\beta}(t)
\end{equation}

 with a self-similarity parameter $H$ given by
\begin{equation}
H =\beta/2+1/\mu-1=[\beta/2-1/2]+[1/\mu]-1/2 \label{eq:H}
\end{equation} more usually known as the Hurst exponent. Note that we would not necessarily expect this equation to hold
for more general fractal processes. In the fBm case $\mu=2$ and
for that case only we recover the well known expression that
$\beta=2H+1$. In the sLm case $\beta=2$ and we find $H=1/\mu$.
Recently \citeauthor{mandelbrot}\shortcite{mandelbrot} has
proposed writing
\begin{equation}
H=J+L-1/2
\end{equation}
where he defines a  Joseph (long range dependence) exponent $J$
($=\beta/2-1/2$) and a Noah (heavy tail) exponent $L$  (=$1/\mu$).


The first property that needs to be shown in  a  time series  for
fLm to be a candidate model is thus $H$-selfsimilarity. This can
be tested by a number of methods. The first is scaling collapse,
which was shown for the datasets in our paper by
\citeauthor{hnat2002b}\shortcite{hnat2002b,hnat2003a}.

An fLm model also implies that the pdf of returns i.e $P(\delta
X=0,\tau)$ will scale with $\tau$ with exponent also equal to $H$.
This was shown in Fig. 2 of \cite{hnat2002b}. For convenience in
figure 1 we show a comparison of the scaling regions of the 1 year
signals taken from the natural time series $AE,AU,AL$ and
$\varepsilon$. All are seen to scale up to approximately $2^6$
minutes ($\approx 1$ hour). Caution is however necessary because
in a natural dataset the moments $|\delta X|^q=|X(t+\tau)-X(t)|^q$
would be expected to be dominated in the small $\tau$ limit by the
scaling of the measurement noise on the differences $\delta X$
rather than that of the physical variables themselves
\cite{hnat2005}.

Interestingly, although the exponent needed to rescale the  pdfs
$P(\delta X,\tau)$ of differences $\delta X$ taken from fLm is the
``full" extended $H=H(\mu,\beta)$   defined in equation
(\ref{eq:H}), the difference pdfs have the same shape  they would
have for an sLm with the same $\mu$ value. This is analogous to
the way in which fBm retains the same Gaussian distribution as the
steps from which it is composed, despite their statistical
dependence, and is why fLm is also known as ``linear fractional
stable motion".  

\begin{figure}[t] \vspace*{2mm}
  \centering{\includegraphics[width=8.3cm]{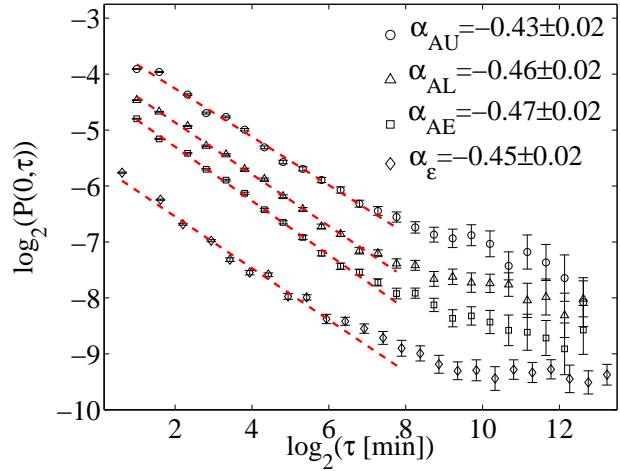}} 
  \caption{ Estimation of  Hurst exponent $H$ via scaling of peaks $P(0)$ of pdfs of differenced time series $X(t+\tau)-X(t)$
 as a function of differencing interval $\tau$. Plots are for i auroral indices
1978:  $X=AU$($\circ$), $X=AL$ ($\triangle$) and AE (box) and ii
solar wind $\varepsilon$ ($\diamond$) for 1995. Plots have been
offset vertically for clarity.} \label{Figure1}
\end{figure}


\subsection{Structure functions $S_q$ and their scaling exponents $\zeta(q)$: $H$  as
$\zeta(1)$, while the pdf of returns gives $\zeta(-1) \equiv -H$}

One may extend the idea  of self-similarity expressed by $H$ to
the generalised q-th order structure functions \cite{frisch}:
\begin{equation}
 S_q = <|x(t+\tau)-x(t)|^{q}>
\end{equation}
where $q$ need not be  integer. If a given $S_q$ is empirically
found to be a power law we can then define an exponent $\zeta(q)$
from $S_q \sim \tau^{\zeta(q)}$.

For a stable self-similar process where all moments are finite
$\mu=2$, i.e. WBm ($H=0.5$)
 or  fBm  ($0 \le H \le 1$), the exponents of the structure functions $\zeta(q)$ follow $\zeta(q)=qH$,
 as we have checked by simulating an  fBm  using the same fLm
  algorithm as used for the figures, in the $\mu=2$ limit.
By definition we then have $\zeta(1)=H$.  Additionally in these
Gaussian ($\mu=2$) cases $\zeta(2)=2H$, which from Equation
(\ref{eq:H}) then implies $2H=\beta-1$.

The exponent derived from the pdf of returns  can be shown to be
equivalent to $\zeta(-1)$ [{\em Miriam Forman, private
communication, 2002}] so for  self-similar processes (see also our
figure 5) the plot of $\zeta(q)$ versus $q$ is antisymmetric about
$q=0$ at least insofar as  $\zeta(-1)=-H=-\zeta(1)$.




\subsection{Second order moment and $J$: Pseudo-Gaussian behaviour of
truncated L\'{e}vy time series}

Because of the relation $\zeta(q)=qH$ for WBm and fBm, a
complementary estimate of the self-similarity parameter $H$ can,
for these  cases, be obtained from  from the well-known growth of
the standard deviation $\sigma(\tau)$ of the difference time
series $\delta X(\tau)$ with differencing interval $\tau$. Indeed
the growth of a measured $\sigma$ as $\tau^{1/2}$ in the case of
WBm defines diffusive behaviour. $\sigma$ is the square root of
variance and thus  scales like $S_2$, i.e. as
$\tau^{(\beta-1)/2}$, i.e. it follows Mandelbrot's
\shortcite{mandelbrot} Joseph exponent $J$ (which from
($\ref{eq:H}$) will be identical to $H$ in the Gaussian WBm or fBm
cases).

\begin{figure}[t]
\vspace*{2mm}
  \centering{\includegraphics[width=8.3cm]{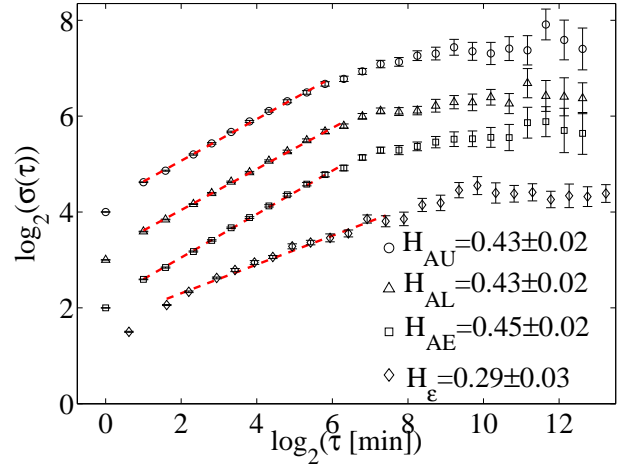}} 
  \caption{ Estimation of  exponent $J$ for scaling of the standard
deviation $\sigma$ of the differenced series versus $\tau$ for the
same quantities as figure 1. Notation as figure 1.
}\label{Figure3}
\end{figure}

In the case of L\'{e}vy motion, however, whether ordinary or
fractional, the $qth$ order moments $S_q$ (where $q > \mu$) taken
from a set of $N$ data points are theoretically infinite as $N
\rightarrow \infty$ in contrast to the convergence seen for
Gaussians. It is thus not {\it a priori} obvious how the variance
of a truncated, finite-$N$, time series would be expected to scale. This is
significant because any simulation that we perform of fractional
L\'{e}vy motion is effectively one of truncated L\'{e}vy motion;
while a natural time series will also have a finite variance in
practice. The possible relevance of this question to data is
clearly illustrated by our Figure (\ref{Figure3}), (see also table
1 of \cite{hnat2002b}) in which $\sigma$ for the solar wind
variable $\varepsilon$ is seen to scale with an exponent of 0.29 as
opposed to the values around $0.43-0.45$ seen for the 3
geomagnetic index quantities. Rather than scaling with $H$,
$\sigma$ still appears to be showing pseudo-Gaussian behaviour
i.e. following  $J$, in that $\beta=1.56$ for this time series
(estimated by wavelet methods) giving $J=(1.56-1)/2=0.28$.


The  apparent disadvantage of the loss of a second, independent,
estimate of $H$ seems to be  compensated for by the possibility
that we can use the growth of $\sigma$ to measure $\beta$ i.e. we
can effectively use it as a measurement of $J$. On the assumption
that a naturally truncated fLm describes our data we can build a table
(Table~\ref{sphericcase}) of the measured $\beta$ and $H$ values
and then predict $\mu$ using equation (\ref{eq:H}).

\begin{table*}
\caption[]{Measured values of $H$ (from Fig. 1) and $J$ (from Fig.
2) for natural time series, and $\mu$ value predicted from
equation (\ref{eq:H}) on the assumption of naturally truncated fLm. All measured
values are $\pm 0.02$ except $J$ for $\varepsilon$ which is $\pm
0.03$.} \label{sphericcase}
\begin{tabular}{crrrrc}
\hline Variable  & \multicolumn{1}{c}{Measured $H$} &
\multicolumn{1}{c}{Measured $J$} & \multicolumn{1}{c}{Inferred
$\beta$} & \multicolumn{1}{c}{Inferred $L$} &
Predicted $\mu$ \\
\hline
$AE$ &   0.47   & 0.45 & 1.90 & 0.52 & 1.92\\
$AU$ &   0.43   & 0.43 & 1.86 & 0.5 & 2  \\
$AL$ &    0.46   & 0.43 & 1.86 & 0.53 & 1.88 \\
$\varepsilon$ &   0.45   & 0.29 & 1.58 & 0.66 & 1.51    \\
 \hline
\end{tabular}
\end{table*}

On inspecting  Table I the first point is that the values of $H$
and $J$ are so close in the case of $AU$ that if we assume they
are exact the predicted $\mu$ becomes 2, eliminating fLm as a
model for $AU$. The $H$ is sub-diffusive, so fBm would remain  a
possible candidate model; however the observed \cite{hnat2003a}
difference pdfs $P(\delta X)$ for $AU$ are non-Gaussian,
eliminating fBm. The error bars quoted in Table I suggest these
conclusions may be too harsh.  fLm would, however, seem more
suitable as a model for $AE,AL$ and $\varepsilon$.

As a test we may also consider the values of $H$ and $J$ for solar
wind $B^2$ obtained by
\citeauthor{hnat2002a}\shortcite{hnat2002a}. Their figure 3 gives
$H=0.42$ in our parlance, while they report a scaling exponent for
$\sigma$ of 0.28 (i.e. $J$). Inserting this into equation
(\ref{eq:H}) predicts $\mu=1.56$, which is equivalent to the
$1/\alpha$ of their equation (3) (see also their Figure 4) which
they find to be $1/0.66=1.5$, encouragingly good agreement.

\subsection{Fractional L\'{e}vy simulation: Comparison with first and second order measures}

We can then now  simulate  fLm using parameters drawn from natural
data to see if the inferences we have drawn above are indeed
consistent, and to qualify fLm as at least a possible proxy for
these time series. We use the published algorithm of
\cite{chicago}. This has the advantage of being linked more
closely to the definition of fLm from equation (\ref{eq:fLm}) than
the (faster) approach of  replacing \cite{chechkin2}a Gaussian
random number generator by a L\'{e}vy generator in otherwise
standard Fourier filter methods \cite{voss1985}. A comparison of
these two approaches will be reported in a future paper.

\begin{figure}[t]
\vspace*{2mm}
  \centering{\includegraphics[width=8.3cm]{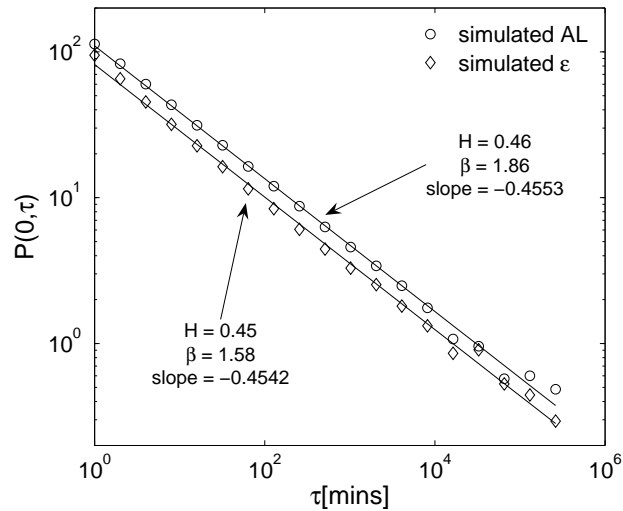}} 
  \caption{ Estimation of  $H$ via scaling of peaks $P(0)$ of pdfs of differenced
  model fractional L\'{e}vy motion time series $X(t+\tau)-X(t)$
 as a function of differencing interval $\tau$. Plots are for i)
 a synthetic $AL$($\circ$) time series and ii) a series of synthetic solar
wind  $X=\varepsilon$ ($\diamond$). Plots have been offset vertically
for clarity. }\label{Figure4}
\end{figure}

\begin{figure}[t]
\vspace*{2mm}
  \centering{\includegraphics[width=8.3cm]{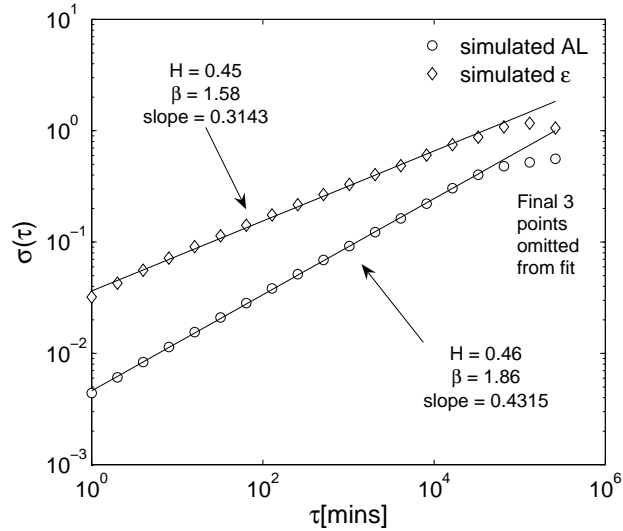}} 
  \caption{ Estimation of $J$ by scaling of the standard
deviation $\sigma$ of the differenced simulated (and thus truncated) series versus $\tau$
for the same quantities as Figure (\ref{Figure3}). Notation as in
(\ref{Figure3}). }\label{Figure5}
\end{figure}

We show simulation results for synthetic $AL$ and $\varepsilon$ time
series. These were specified by the ordered pairs $(\beta,\mu)$ of
(1.86,1.88) and (1.58,1.51) respectively. The $P(\delta X=0,\tau)$
scaling for both series (Figure \ref{Figure4}) is seen to follow
$H$ as we expect, so both model series have very similar measured
$H$ values, as we also  saw in their natural counterparts (Figure
(\ref{Figure1})). Conversely, for finite samples of fLm, however,
modelling $AL$ and $\varepsilon$ we see from Figure (\ref{Figure5})
that rather than following $\tau^{1/\mu} (=\tau^L)$, the $\sigma$
measured on the difference time series $\delta X$ still grows as
$\tau^{(\beta-1)/2} (=\tau^{J})$ i.e. it does, as postulated in
subsection  3.5, measure $J$ rather than $L$.

This effect requires some discussion. It seems to be a further
manifestation of the ``pseudo-Gaussian" behaviour of truncated
standard L\'{e}vy motion \cite{chechkin1}, and known \cite{nakao}
to be responsible for the result $\zeta(2)=1$ in that case (see
also Figure 5). Our simulations have clearly demonstrated that it
generalises to the long-range dependent fLm case i.e. that in
general for fLm $\zeta(2)/2=J=(\beta-1)/2$. This conclusion is
most clearly supported by  Figure 5 where the $\zeta(2)$ value can
be read off as following this relation over the range $\beta=1.5$
to $2.5$. The agreement is poorer at smaller $\beta$ values
tested. We currently think this reflects known difficulties with
accurately simulating strongly anticorrelated fLm
\cite{chechkin2}. The effect has previously been remarked on in
the truncated standard L\'{e}vy  paradigm; for example the S\&P
500 financial time series, depicted by \cite{MantegnaStanley2000}
where $\beta=2$ (their Fig. 11.4.a) so $\sigma$ grows as
$\tau^{1/2}$ (their Fig. 11.3a), in contrast to an $H$ value from
peak scaling of $0.71$ (their Fig. 9.3).

In the multifractal modelling community the power spectrum has
long been seen as only just one of several ways of measuring
$\zeta(2)$. 
For this reason a difference in the
value of $\zeta(1) \ne \zeta(2)/2$ has sometimes been claimed as
direct evidence of the inapplicability of {\it any} additive model
and thus the immediate need for a multiplicative model
\cite{SchertzerLovejoy1987}.

Conversely our result would seem to suggest that any truncated
stable additive model other than the fBm/WBm limiting cases is
likely to show $\zeta(1) \ne \zeta(2)/2$, and
$\zeta(2)/2=J=(\beta-1)/2$ without the need for  a multiplicative
model. This may be understood as being because truncated L\'{e}vy
motion, whether standard or fractional, behaves as a bifractal
\cite{nakao}. There may be natural time series where additive fLm
is actually the most natural model, or at least an economical and
easily specified one.


\begin{figure}[t]
\vspace*{2mm}
  \centering{\includegraphics[width=8.3cm]{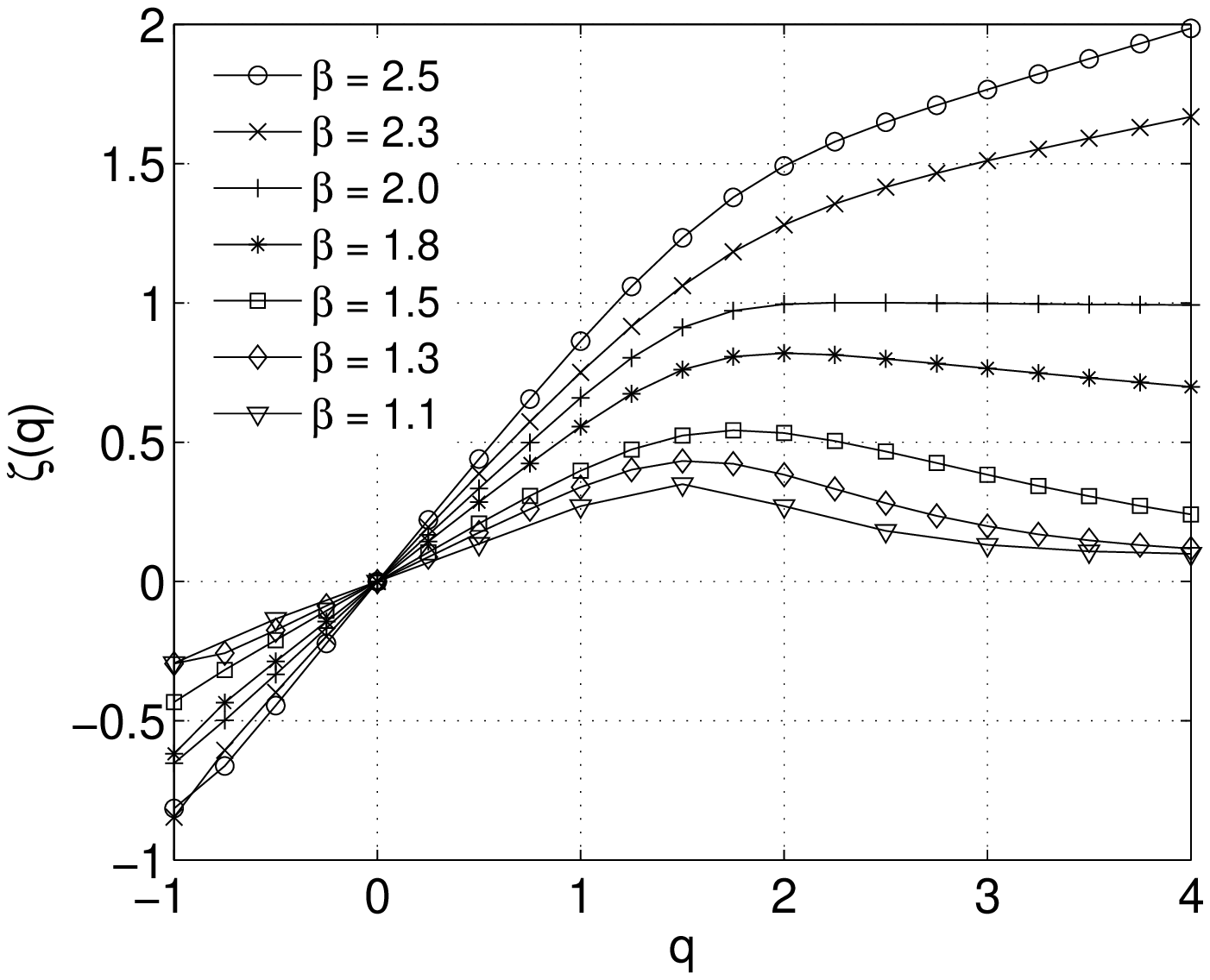}} 
  \caption{ Zeta plots for simulated (and thus truncated) fLm with $\mu$ fixed at 1.5
  and $\beta$ ranging from 1.1 to 2.5. The relation $\zeta(2)=\beta-1$ is seen to be well
  satisfied for $\beta \ge 1.5$.
}\label{Figure7}
\end{figure}


\subsection{$\zeta$ plots and the multifractality of truncated L\'{e}vy motions}

\begin{figure}[t]
\vspace*{2mm}
  \centering{\includegraphics[width=8.3cm]{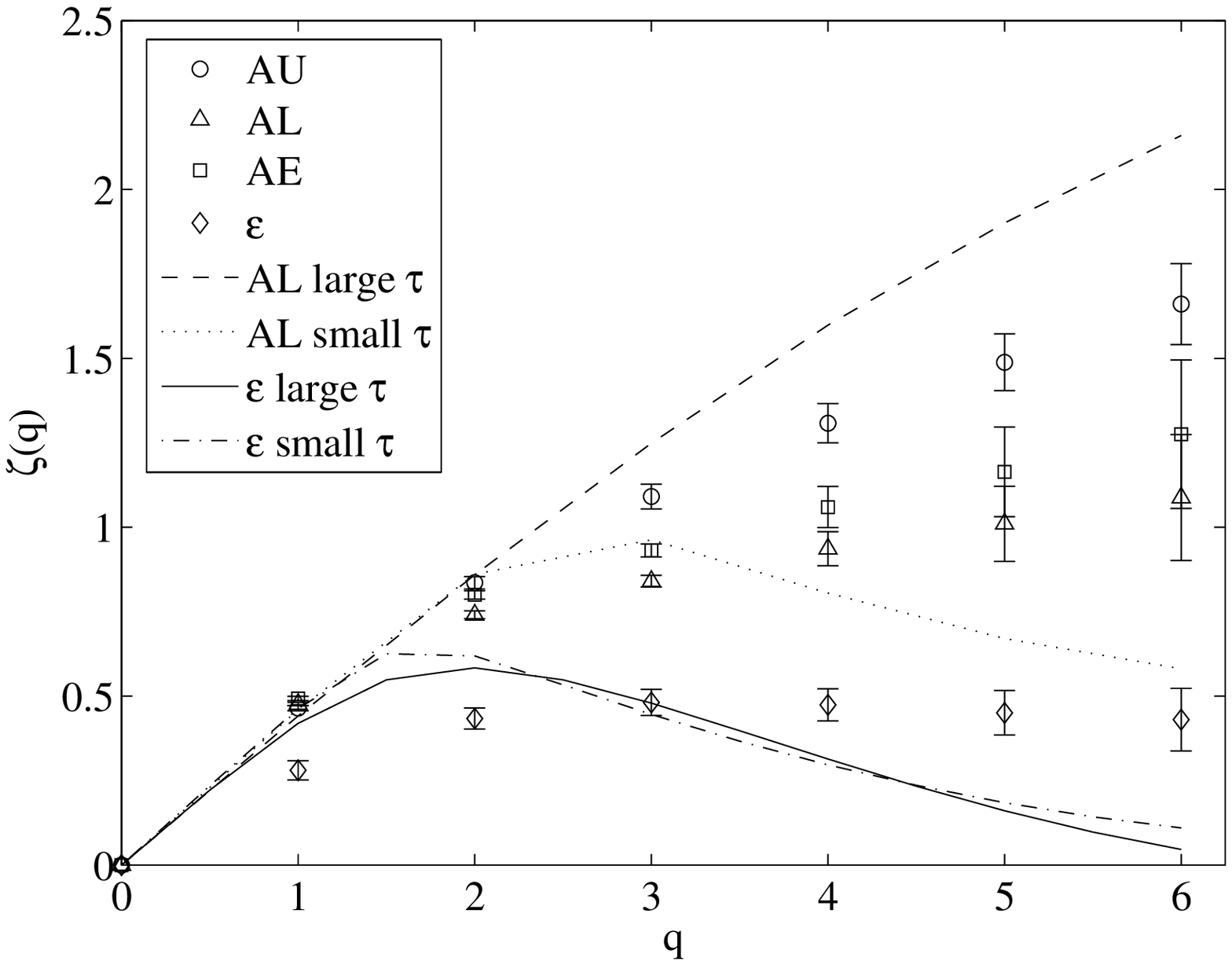}} 
  \caption{ $\zeta(q)$ versus $q$ plots for  3 auroral indices ($AU,AL$ and $AE$) during 1978, and
  solar wind $\varepsilon$ from WIND for 1985. Overlaid are $\zeta(q)$ plots for our simulated
  AL, and  simulated $\varepsilon$, where in both cases ``large $\tau$" and ``small $\tau$" indicate
  the range of estimates depending on the part of the structure function taken to be
  scaling.}
\label{Figure8}
\end{figure}

At this point it may be objected that we have not tested any
predictions of the fLm model against the behaviour of natural time
series other than those properties used to specify it. Our first
additional check is thus to examine the multi-affine behaviour
seen in the data and the model using the ``$\zeta$ plots" defined
in section 3.4. Such a plot, showing scaling exponent $\zeta(q)$
versus moment $q$ is shown for the data in figure (\ref{Figure8}).
Interestingly $AU$ most resembles a ``classic" multifractal, in
that the points $\zeta(q)$ lie on a curve rather than a straight
or broken line \cite{frisch}. However $AE$, or at least $AL$, have
$\zeta$ which arguably flattens out near 1 for higher moments.
$\varepsilon$ intriguingly even seems to {\it fall} as $m$ increases.
This behavior is qualitatively similar to that seen for our
simulated $AL$ and $\varepsilon$  time series, whose $\zeta(q)$ plots
are superposed on the figure. In particular a change in the range
of $\tau$ over which the simulated $AL$ structure functions are
taken to be power laws is enough to encompass the observed range
of $\zeta$ plots for $\varepsilon$.  More detailed comparison is at
present prevented by the difficulty of obtaining accurate values
of $S_q$ for high moments-an issue also afflicting analysis of
real data.

\section{Conclusions}
A significant body of data and models now exists for the problem
of solar wind and magnetic index scaling. We have here suggested a
complementary approach, motivated in particular by the need to i)
reconcile differing estimates of scaling exponents (in hindsight
the Joseph and Hurst exponents $J$ and $H$); ii) model
subdiffusive behaviour ($H < 0.5$); and iii) model long-ranged
correlation ($\beta \ne 2$). We proposed the use of a simple and
economical model: fractional L\'{e}vy motion, to describe the
scaling of the above quantities. Initial consistency checks with
respect to the distribution of returns and the scaling of standard
deviation support the use of fLm, and the multi-affine ``zeta
plots" are more qualitatively similar. Importantly we find that
the degree of similarity between model solar wind $\varepsilon$ and
the model $AL$ index does indeed depend on the moment order at
which comparison is made, but that this does not, however, require
a multiplicative process to explain it. The difference can,
rather, be understood as coming from the bifractality of a
truncated fractional L\'{e}vy motion. This explains why some
measures such as  $H$  from the distribution of returns or pdf
rescaling are much closer to each other than, for example, the
$\sigma$-based exponent (which we found to measure $J$, not $H$).

The present paper has been mainly concerned with the modelling of measured quantities rather than the 
extent to which  they are artificial.  For geomagnetic indices and other constructed quantitities like $\varepsilon$, however, the extent to which  scaling behaviour could be an artefact of the construction method  is an  important issue. We are aware of some  progress in studying this problem  
(e.g. \citeauthor{edwards} \shortcite{edwards}, \citeauthor{weigel} \shortcite{weigel}), more will be needed.  Further work is also underway to test the predictions of the fLm model against other scaling studies such as the cited burst
lifetime and spreading exponent investigations.

\vspace{-.35in}

\acknowledgements We acknowledge the provision of data by WIND SWE
and MFI teams, and the World Data Centre at RAL. We are grateful
to Gary Abel, Miriam Forman, Sean Lovejoy, Murray Parkinson,
George Rowlands, Misha Sitnov, Zoltan V\"{o}r\"{o}s and James
Wanliss for many helpful interactions.

\end{article}

\begin{thebibliography}{}



\bibitem[\protect\citeauthoryear{Bruno et~al.} {2004}]{bruno2004} Bruno, R.,
L. Sorriso-Valvo, V. Carbone, and B. Bavassano. \newblock {A
Possible Truncated-L\'{e}vy-flight Statistics Recovered From
Interplanetary Solar-wind
 Velocity and Magnetic-field Fluctuations}.
\newblock {\em Europhysics Letters}, 66(1):146--152, 2004.




\bibitem[\protect\citeauthoryear{Chang  and Consolini}{2001}]{chang2001}
Chang, T.~S., and G. Consolini. \newblock{Magnetic Field Topology
and Criticality in Geotail Dynamics: Relevance to Substorm
Phenomena}.
\newblock {\em Space Science Reviews}, 95(1--2):309-321, 2001.


\bibitem[\protect\citeauthoryear{Chapman et~al.}{2005}]{chapman2005} Chapman, S.~C., B. Hnat, G.
Rowlands, and N.~W. Watkins. \newblock{Scaling Collapse and
Structure Functions: Identifying Self-Affinity in Finite Length
Time Series}. \newblock{\em Nonlinear Processes in Geophysics},
12:767-774, 2005.

\bibitem[\protect\citeauthoryear{Chechkin and Gonchar}{2000a}]{chechkin1}
Chechkin, A. V., and V. Yu. Gonchar.
\newblock{Self and Spurious Multi-affinity of Ordinary L\'{e}vy Motion,
 and Pseudo-Gaussian Relations}. \newblock{\em Chaos, Solitons and
 Fractals}, 11(14):2379-2390, 2000a.

\bibitem[\protect\citeauthoryear{Chechkin and Gonchar}{2000b}]{chechkin2} Chechkin, A. V., and V. Yu. Gonchar.
\newblock{A Model for Persistent L\'{e}vy Motion}.
\newblock{\em Physica A}, 277:312--326, 2000b.

\bibitem[\protect\citeauthoryear{Consolini et~al.}{1996}]{consolini96}
Consolini, G., M. F. Marcucci, and M. Candidi.
\newblock{Multifractal Structure of Auroral Electrojet Index
Data}. \newblock{\em Physical Review Letters}, 76:4082--4085,
1996.


\bibitem[\protect\citeauthoryear{Consolini et al.}{1997}]{consolini1997b} Consolini, G.,
L. Cafarela, P. De Michelis, M. Candidi, and A. Meloni
\newblock{Non-Gaussian Probability Distribution of Short Time Scale Magnetic Field Fluctuations at Terra Nova Bay
(Antarctica)}.
\newblock In S.~Aiello, N. Iucci, G. Sironi, A. Treves and U. Villante, editors,
{\em  Cosmic Physics in the Year 2000},  SIF Conference
Proceedings Volume 58. SIF, Bologna, Italy, 1997.

\bibitem[\protect\citeauthoryear{Davis and Sugiura}{1966}]{davis}
Davis, T.~N., and M. Sugiura. \newblock{ Auroral Electrojet
Activity Index {\it AE} and its Universal Time Variations}.
\newblock{\em Journal of Geophysical Research}, 71:785--801, 1966.

\bibitem[\protect\citeauthoryear{Edwards}{2001}]{edwards}
Edwards, J.~W., A.~S. Sharma, and M. I. Sitnov. \newblock{Spatio-temporal
Structure of Geomagnetic Activity Triggered by Dynamic Pressure Pulses: Mutual Information Functional Analysis}.
\newblock{\em  Bulletin of the American Physical Society}, 27:156, 2001.




\bibitem[\protect\citeauthoryear{Freeman et~al.}{2000a}]{freeman2000a} Freeman,
M. P., N.W. Watkins, and D.J. Riley.
 \newblock{Evidence for a Solar Wind Origin of the Power law Burst
 Lifetime Distribution of the AE Indices}.
 \newblock{\em Geophysical Research Letters}, 27:1087--1090, 2000a.

\bibitem[\protect\citeauthoryear{Freeman et~al.}{2000b}]{freeman2000b} Freeman,
M. P., N.W. Watkins, and D.J. Riley. \newblock{Power law
Distributions of Burst Duration and Interburst Interval in the
Solar Wind: Turbulence or Dissipative Self-organized
Criticality?}.\newblock{\em Physical Review E}, 62(6):8794-8797,
2000b.


\bibitem[\protect\citeauthoryear{Frisch}{1995}]{frisch} Frisch,
U. \newblock{\em Turbulence: the Legacy of A. N. Kolmogorov}.
\newblock Cambridge University Press, 1995.

\bibitem[\protect\citeauthoryear{Hnat et~al.}{2002a}]{hnat2002a} Hnat, B., S.~C. Chapman,
G.~Rowlands, N.~W. Watkins, and W.~M. Farrell. \newblock{Finite
Size Scaling in the Solar Wind Magnetic Field Energy Density as
Seen by WIND}. \newblock{\em Geophysical Research Letters},
29(10), doi:10.1029/2001GL014587, 2002a.

\bibitem[\protect\citeauthoryear{Hnat et~al.}{2002b}]{hnat2002b} Hnat, B., S.~C. Chapman, G. Rowlands,
N.~W. Watkins, and M.~P. Freeman. \newblock{Scaling of Solar Wind
$\varepsilon$ and the $AU,AL$ and $AE$ Indices as Seen by WIND}.
\newblock{\em Geophysical Research Letters}, 29(22), 2078, doi:10.1029/2002GL016054, 2002b.

\bibitem[\protect\citeauthoryear{Hnat et~al.}{2003a}]{hnat2003a} Hnat, B., S.~C. Chapman, G. Rowlands,
N.~W. Watkins, M.~P. Freeman. \newblock{Correction to ``Scaling of
Solar Wind $\varepsilon$ and the $AU,AL$ and $AE$ Indices as Seen by
WIND"}. \newblock{\em Geophysical Research Letters}, 30:(8), 1426,
doi:10.1029/2003GL017194, 2003a.


\bibitem[\protect\citeauthoryear{Hnat et~al.}{2003b}]{hnat2003b} Hnat, B., S.~C. Chapman, G.
Rowlands, N.~W. Watkins, and M.~P. Freeman. \newblock{Scaling in
Long Term Data Sets of Geomagnetic Indices and Solar Wind Epsilon
as Seen by WIND Spacecraft}. \newblock{\em Geophysical Research
Letters}, 30(22), 2174, doi:10.1029/2003GL018209, 2003b.

\bibitem[\protect\citeauthoryear{Hnat et~al.}{2005}]{hnat2005} Hnat, B., S.~C. Chapman, and G.
Rowlands, \newblock{Scaling and a Fokker-Planck model for
Fluctuations in Geomagnetic Indices and Comparison with Solar Wind
$\varepsilon$ as Seen by WIND and ACE}. {\em Journal of Geophysical
Research}, 110, A08206, doi:10.1029/2004JA010824, 2005. 

\bibitem[\protect\citeauthoryear{Kabin and
Papitashvili}{1998}]{kabin1998} Kabin, K., and V.~O. Papitashvili.
\newblock{Fractal Properties of the IMF and the Earth's Magnetotail Field}.
\newblock{\em Earth Planets Space}, 50:87-90 (1998).



\bibitem[\protect\citeauthoryear{Laskin et~al.}{2002}]{laskin2002} Laskin, N., I. Lambadaris, F.
Harmantzis, and M. Devetsikiotis. \newblock{Fractional L\'{e}vy
motion and its application to network traffic modelling}.
\newblock{\em Computer Networks}, 363-375 (2002).




\bibitem[\protect\citeauthoryear{Mandelbrot}{1995}]{mandelbrot1995} Mandelbrot, B.~B.
\newblock{Introduction to Fractal Sums of Pulses}. \newblock In M.~F. Shlesinger,
G.~M. Zaslavsky, and U. Frisch, editors, {\em L\'{e}vy flights and
Related Topics in Physics: Proceedings of the International
Workshop, Nice, France, June, 1994}. Lecture Notes in Physics:
450. Springer-Verlag, Berlin, 1995.

\bibitem[\protect\citeauthoryear{Mandelbrot}{2002}]{mandelbrot} Mandelbrot, B.~B.
\newblock{\em Gaussian Self-Affinity and Fractals: Globality, the Earth, 1/f Noise and R/S  }. Springer Verlag, 2002.


\bibitem[\protect\citeauthoryear{Mantegna and Stanley}
{2000}]{MantegnaStanley2000} Mantegna, R.~N., and H.~E. Stanley.
\newblock {\em  An Introduction to Econophysics: Correlations and Complexity
in Finance}. \newblock Cambridge University Press, 2000.

\bibitem[\protect\citeauthoryear{March et al}{2005}]{march2005} March, T.~K., S.~C. Chapman, and
R. O. Dendy. \newblock{Mutual Information Between Geomagnetic
Indices and the Solar Wind as Seen by WIND: Implications for
Propagation Time Estimates}. \newblock{\em Geophysical Research
Letters}, 32, L04101,doi:10.1029/2004GL021677, 2005

\bibitem[\protect\citeauthoryear{Milovanov and Zelenyi}{2001}]{milovanov2001} Milovanov, A.~V.,
and L.~M. Zelenyi. \newblock {``Strange" Fermi Processes and Power-law Nonthermal Tails From a 
Self-consistent Fractional Kinetic Equations}. \newblock {\em Physical Review E}, 
64:052101, 2001.

\bibitem[\protect\citeauthoryear{Nakao}{2000}]{nakao} Nakao, H.
\newblock{Multi-scaling Properties of Truncated L\'{e}vy Flights}.
\newblock{\em Physics Letters A}, 266(4--6), 282-289, 2000.


\bibitem[\protect\citeauthoryear{Painter and Patterson}{1994}]{painter}
Painter, S., and L. Patterson. \newblock{Fractional L\'{e}vy
Motion as a Model for Spatial Variability in Sedimentary rock}.
\newblock{\em Geophysical Research Letters}, 21(25): 2857-2860,
1994.

\bibitem[\protect\citeauthoryear{Perreault and Akasofu}{1978}]{perrault}
Perreault, P., and S.-I. Akasofu. \newblock{A Study of Geomagnetic
Storms}.  \newblock{\em Geophysical Journal of the Royal
Astronomical Society}, 54: 547--573, 1978.



\bibitem[\protect\citeauthoryear{Schertzer and Lovejoy}{1987}]{SchertzerLovejoy1987}
Schertzer, D., and S. Lovejoy. \newblock{Physical Modeling and
Analysis of Rain and Clouds by Anisotropic Scaling Multiplicative
Processes}. \newblock{\em Journal of Geophysical Research},
92(D8):9693--9714, 1987.



\bibitem[\protect\citeauthoryear{Takalo et~al.}{1993}]{TakaloGRL}
Takalo, J., Timonen, J., and H. Koskinen.\newblock{Correlation
Dimension and Affinity of AE Data and Bicolored Noise}.
\newblock{\em Geophysical Research Letters}, 20(15): 1527-1530,
1993.


\bibitem[\protect\citeauthoryear{Tsurutani et~al.}{1990}]{tsurutani}
Tsurutani, B.~T., M. Sugiura, T. Iyemori, B.~E. Goldstein, W.~D.
Gonzalez, S.-I. Akasofu, E.~J. Smith. \newblock{The Nonlinear
Response of AE to the IMF $B_s$ Driver: A Spectral Break at $5$
Hours}. \newblock{\em Geophysical Research Letters}, 17:279--282,
1990.


\bibitem[\protect\citeauthoryear{Ukhorskiy et~al.}{2004}]{ukhorskiy2004} Ukhorskiy, A.~Y., Sitnov, M.~I., Sharma A.~S.,
Papadopoulos K. \newblock{Global and Multi-scale Features of Solar
Wind-magnetosphere Coupling: From Modeling to Forecasting,  {\em
Geophysical Research Letters}, 31(8):L08802, 2004.


\bibitem[\protect\citeauthoryear{Uritsky et~al.}{2001}]{uritsky} Uritsky,  V.~M.,
 A.~J. Klimas and D. Vassiliadis. \newblock{Comparative Study of Dynamical Critical
 Scaling in the Auroral Electrojet Index Versus Solar Wind
 Fluctuations}. \newblock{\em Geophysical Research Letters}, 28:3809--3812, 2001.

\bibitem[\protect\citeauthoryear{V\"{o}r\"{o}s et~al.}{1998}]{voros} V\"{o}r\"{o}s, Z., P. Kovacs, A. Juhasz,
A. Kormendi and A.~W. Green. \newblock{Scaling Laws from
Geomagnetic Time Series}. \newblock{\em Geophysical Research
Letters} 25:2621--2624, 1998.

\bibitem[\protect\citeauthoryear{Voss}{1985}]{voss1985} Voss, R.~F.
\newblock Fractals in Nature: From Characterization to Simulation.
\newblock In Heinz-Otto Peitgen and Dietmar Saupe, editors,
{\em The Science of Fractal Images}. Springer-Verlag, Berlin
  Heidelberg New York Tokyo, 1985

\bibitem[\protect\citeauthoryear{Watkins}{2002}]{WatkinsNPG}
Watkins, N. W. \newblock{Scaling in the Space Climatology of the
Auroral Indices: is SOC the Only Possible Description?}.
\newblock{\em Nonlinear Processes in Geophysics}, 9(5-6):389--397,
2002.

\bibitem[\protect\citeauthoryear{Weigel and Baker}{2003}]{weigel} Weigel, R.~S. and D.~N. Baker. \newblock{Probability distribution invariance of 1-minute auroral zone 
geomagnetic field fluctuations}.
\newblock{\em Geophysical Research Letters}, 30(23), 2193, doi:10.1029/2003GL018470, 2003.


\bibitem[\protect\citeauthoryear{Wu et~al.}{2004}]{chicago} Wu, W.~B.,
G. Michailidis, and D. Zhang. \newblock{Simulating Sample Paths of
Linear Fractional Stable Motion}. \newblock{\em IEEE Transactions
On Information Theory}, 50(6):1086--1096, 2004.


\bibitem[\protect\citeauthoryear{Zaslavsky}{2002}]{zaslavsky} Zaslavsky, G.~M.,
\newblock{Chaos, Fractional Kinetics and Anomalous Transport}. \newblock{\em Physics Reports}, 371:461-580, 2002.

}

\end{thebibliography}
\end{document}